\documentclass[letterpaper]{emulateapj}
\usepackage{amsmath}
\usepackage[varg]{txfonts}

\let\url\relax
\usepackage[hypertex]{hyperref}
\usepackage[all]{hypcap}
\makeatletter
\renewcommand{\fps@figure}{tp}
\makeatother
\setkeys{Gin}{width=\linewidth,height=0.9\textheight,keepaspectratio=true}

\newcommand{\pcc}{\ensuremath{\mathrm{cm^{-3}}}}
\newcommand{\kms}{\ensuremath{\mathrm{km~s^{-1}}}}
\newcommand{\Hii}{\ion{H}{2}}
\newcounter{ionstage}
\renewcommand{\ion}[2]{\setcounter{ionstage}{#2}%
  \ensuremath{\mathrm{#1\,\scalebox{0.9}[0.8]{\Roman{ionstage}}}}}

\begin{document}

\shorttitle{\Hii{} regions in turbulent clouds}
\shortauthors{Mellema, Arthur, Henney, Iliev, \& Shapiro}

\title{Dynamical \Hii{} Region Evolution in Turbulent
  Molecular Clouds}

\author{Garrelt Mellema\altaffilmark{1,2}, S. Jane
  Arthur\altaffilmark{3}, William J. Henney\altaffilmark{3}, Ilian
  T. Iliev\altaffilmark{4}, and Paul R. Shapiro\altaffilmark{5}}

\altaffiltext{1}{ASTRON, P.O. Box 1, NL-7990 AA Dwingeloo, The
  Netherlands; {gmellema@astron.nl} }

\altaffiltext{2}{Sterrewacht Leiden, P.O. Box 9513, NL-2300 RA Leiden,
  The Netherlands.}  

\altaffiltext{3}{Centro de Radioastronom\'{\i}a y Astrof\'{\i}sica,
  UNAM, Campus Morelia, Apartado Postal 3-72, 58090 Morelia, M\'exico;
  {j.arthur@astrosmo.unam.mx, w.henney@astrosmo.unam.mx} }

\altaffiltext{4}{Canadian Institute for Theoretical Astrophysics,
  University of Toronto, 60 St.\ George Street, Toronto, ON M5S 3H8,
  Canada; {iliev@cita.utoronto.ca}}

\altaffiltext{5}{Department of Astronomy, The University of Texas at
  Austin, RLM 16.204, Austin, TX 78712; {shapiro@astro.as.utexas.edu}}

\begin{abstract}
  We present numerical radiation-hydrodynamic simulations of the
  evolution of \ion{H}{2} regions formed in an inhomogeneous medium
  resulting from turbulence simulations. We find that the filamentary
  structure of the underlying density distribution produces a highly
  irregular shape for the ionized region, in which the ionization
  front escapes to large distances in some directions within
  80,000~years. In other directions, on the other hand, neutral gas in
  the form of dense globules persists within 1~parsec of the central
  star for the full duration of our simulation
  (400,000~years). Divergent photoablation flows from these globules
  maintain a root-mean-squared velocity in the ionized gas that is
  close to the ionized sound speed. Simulated images in optical
  emission lines show morphologies that are in strikingly detailed
  agreement with those observed in real \Hii{} regions.
\end{abstract}

\keywords{\ion{H}{2} regions --- ISM: clouds --- ISM: kinematics and
  dynamics --- stars:formation --- turbulence}

\section{Introduction}
\label{sec:Intro}

The currently accepted view of the interstellar medium (ISM) is that
its density and velocity distributions are shaped by turbulence
\citep[see, e.g.][]{{2000prpl.conf....3V}, {2004ARA&A..42..211E}}. In
the molecular clouds that will constitute the sites of star formation,
the ISM turbulence is supersonic, with turbulent Mach numbers as large
as 10 or more \citep{1981MNRAS.194..809L}. Such high Mach numbers
indicate that the medium is highly compressible.  The efficiency of
the star formation process is known to be small from observations, at
most 10\%--30\% for cluster-forming cores
\citep{2003ARA&A..41...57L}. In numerical simulations of supersonic
turbulent media, large-amplitude density fluctuations are generated,
which may become locally gravitationally unstable and collapse to form
stars.  

Once a massive star forms, it will photoionize its surroundings,
forming an \ion{H}{2} region. Observed \ion{H}{2} regions display a
wide variety of morphologies and sizes and are generally
irregular. While still embedded in their natal molecular clouds,
\ion{H}{2} regions remain small and are classified as ultracompact
(linear size $< 0.1$~pc) or compact (0.1--1~pc)
\citep{1989ApJS...69..831W}. Extended \ion{H}{2} regions (size $>
1$~pc) are thought to correspond to more evolved states. The formation
of \ion{H}{2} regions and the propagation of ionization fronts in
uniform media has been well studied
\citep{{1939ApJ....89..526S},{1954BAN....12..187K}}. The expansion of
\ion{H}{2} regions in power-law and plane-parallel density
distributions, and the breakout of ionized gas from dense clouds into
a low density medium has also been investigated
\citep{{1979A&A....71...59T},
  {1981A&A....98...85B},{1989RMxAA..18...65F},
  {1990ApJ...349..126F},{2005ApJ...627..813H},{2005astro.ph.11035A}}.
These models all assume a smooth density variation and do not give
rise to the irregular complex shapes exhibited by real \ion{H}{2}
regions \citep{2006-FS-Will}.

It has been suggested that the irregular shapes of observed \ion{H}{2}
regions, including the ``fingers'' and ``elephant trunks'' seen in
regions such as M16 \citep[see, e.g.,][]{1996AJ....111.2349H}, are due
to instabilities in the ionization front. When cooling is included in
radiation-hydrodynamical simulations of \ion{H}{2} region expansion, a
strong instability, predicted analytically by
\citet{1979ApJ...233..280G}, leads to vigorous fragmentation of the
expanding massive shell \citep{1996ApJ...469..171G}. When small-scale
density fluctuations are present in the ambient medium,
\citet{1999MNRAS.310..789W} shows that a shadowing instability in the
R-type ionization front phase can lead to a non-linear hydrodynamic
instability and the formation of clumps in the D-type phase. The
clumps and filaments formed by such instabilities tend to be radial in
nature, and thus do not account for observed structures such as the
Orion bar, which has a linear appearance, perpendicular to the
direction to the ionizing star.

It is entirely possible that the irregular appearance of observed
\ion{H}{2} regions is due not to instabilities in the ionization front
but to underlying structure in the ambient
medium. \citet{2004ApJ...610..339L} calculated the propagation of the
R-type ionization front produced by a massive star in a precomputed
three-dimensional compressible turbulent density field. Their results
show that the initial density structure is important for the resulting
structure of the \ion{H}{2} region. However, these calculations do not
include hydrodynamics and therefore cannot follow the subsequent
dynamical expansion of the ionized gas into the surrounding turbulent
medium after the initial $\sim 100$~yrs. More recently,
\citet{2005MNRAS.358..291D} have presented
smooth-particle-hydrodynamic simulations of the development of an
\Hii{} region around a newly-formed star cluster. These simulations do
follow the dynamical expansion of the ionized gas in a highly clumped
medium and its feedback on the star formation process, but the initial
density field is specified in a somewhat ad~hoc manner.

In this article we present results from the first fully coupled,
radiation-hydrodynamic simulations of the formation and expansion of
\ion{H}{2} regions in a turbulent medium. As our starting point, we
take a precomputed three-dimensional compressible turbulent density
field calculated by \citet{2005ApJ...618..344V}. In \S~\ref{sec:simul}
we describe our numerical method and the initial conditions. Our
results are presented in \S~\ref{sec:results} in a form directly
comparable to observations. In \S~\ref{sec:discuss} we summarize and
conclude our findings.

\section{Simulations}
\label{sec:simul}
\subsection{Numerical Method}
\label{subsec:method}
The hydrodynamics is calculated using the non-relativistic scheme
described in \citet{1995A&AS..110..587E}, with the addition of a Local
Oscillation Filter \citep{2003ApJS..147..187S} to suppress numerical
odd-even decoupling behind radiatively cooling shock waves. For the
radiative transfer, we implement the new, efficient C$^2$-Ray method,
described by \citet{2005astro.ph..8416M}, which is explicitly photon
conserving. The C$^2$-Ray algorithm uses an analytical relaxation
solution for the ionization rate equations, which allows for time
steps much larger than the characteristic ionization timescales and
the timescale for the ionization front to cross a numerical grid
cell. This efficiency allowed us, in another current application, to
use this code to perform the first large-scale simulations of the
reionization of the universe, which required solving for the radiative
feedback of up to tens of thousands of sources on 200$^3$--400$^3$
computational grids \citep{reion_sim}.  The radiative transfer is
coupled to the hydrodynamics via operator splitting and tests of the
combined numerical method are presented in \citet{2006astro.ph..3199I,
  2006CodeComparison}.\footnote{Details of the test problems can be
  found at \url{http://www.mpa-garching.mpg.de/tsu3/}.} These features
make the current problem of three-dimensional \ion{H}{2} region
expansion in a turbulent medium entirely tractable, and reduce the
computational time for the simulations to a few hours for a $128^3$
grid on a single processor. The simulations presented in this paper
are for a single, fixed grid since the radiative transfer algorithm is
parallelized only for shared-memory machines.

\subsection{Initial Conditions}
\label{subsec:initcond}
As initial conditions for the density and velocity fields we take the
three-dimensional numerical simulations of driven turbulence presented
by \citet{2005ApJ...618..344V}. These turbulence models are isothermal
but once included in our simulations the gas is subject to heating due
to photoionization, and cooling, calculated as in
\citet{1999RMxAA..35..123R}. We use the simulations with
root-mean-squared (rms) sonic Mach number $M_s = 10$, which give a
large dynamic range for the initial density
fluctuations. \citet{2005ApJ...618..344V} present both
ideal-magnetohydrodynamics (MHD) and nonmagnetic cases but we have
chosen a nonmagnetic simulation for the present paper because our
hydrodynamics scheme does not include MHD. The turbulence simulations
are scale free and are characterized by three nondimensional numbers:
$M_s = \sigma/c$ (the rms sonic Mach number, where $\sigma$ is the
turbulent velocity dispersion and $c$ is the sound speed), $J \equiv
L/L_J$ (the Jeans number, giving the size of the box in units of the
Jeans length $L_J$), and $\beta \equiv P_{\rm th}/P_{\rm mag}$ (the
ratio of thermal to magnetic pressures). In the case we consider, $M_s
= 10$, $J = 4$, and $\beta = \infty$. The turbulence calculation is
time dependent and we choose a late time, when several regions are
collapsing, as the initial state for the principal run of our
radiation-hydrodynamics simulation. 

The scaling to physical variables we choose is that used by
\citet{2005ApJ...618..344V}, that is, isothermal sound speed $c_s =
0.2$~km~s$^{-1}$ and mean nucleon number density\footnote{The original
  simulation used molecular gas with mean molecular mass $2.4
  m_\mathrm{H}$ and number density 500~\pcc. Since we do not treat the
  dissociation of molecular hydrogen in our code, we have substituted
  atomic hydrogen with the same mass density and sound speed. Another
  consequence of this change is that our initial temperatures are
  lower, but this is of no dynamical consequence.} $n_0 =
932$~cm$^{-3}$. This gives a spatial size $L = 4$~pc for the
computational grid.  We place the ionizing photon source at the center
of the densest clump, which has density $n \simeq 1000 n_0$, and the
periodic boundary conditions used for the turbulence simulation enable
us to move this position to the center of the computational grid.  We
found that the densest clump in the turbulence simulations has a large
absolute velocity, since it forms due to velocity fluctuation in the
molecular gas. Consequently, it is necessary to shift the velocities
of the whole grid by an amount equal to the velocity of the densest
cell, in order that the densest material remains at rest in the frame
of the star. For the ionizing source, we choose a stellar ionizing
photon rate $S_* = 5 \times 10^{48}$~s$^{-1}$ and effective
temperature $T_{\rm eff} = 37500$~K, equivalent to an O7.5 main
sequence star \citep{1973AJ.....78..929P}.  In order to account for
the material that goes into the formation of the ionizing star, we
remove 90\% of the mass from the central $2\times 2\times2$ group of
cells, which corresponds to $\simeq 27 M_\odot$.

Although the initial turbulence simulation has periodic boundary
conditions, our hydrodynamics calculation has open boundaries. This is
because in cases where the photoionized region expands beyond the grid
boundary, periodic boundary conditions would be
unphysical. Furthermore, our calculations do not include self-gravity
and our treatment of the heating and cooling of the cold
neutral/molecular gas is only approximate and does not include
chemistry.

\section{Results}
\label{sec:results}
\begin{figure*}
  \centering
  \includegraphics{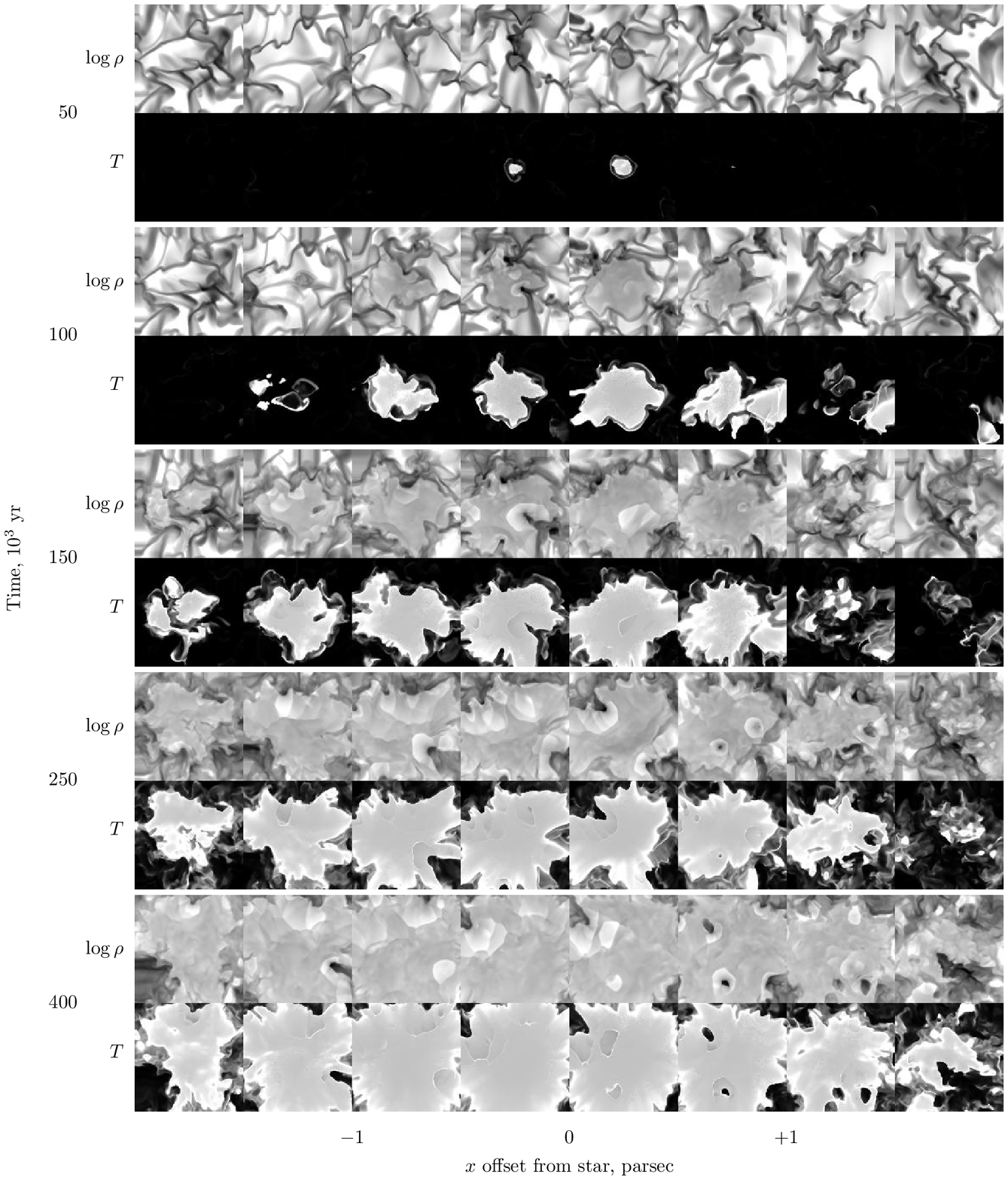}
  \caption{Gas density and temperature slices through our simulation
    for a sequence of evolutionary times, increasing from top to
    bottom, as indicated on the left axis. Density is shown on a
    negative logarithmic scale between $10$ (white) and
    $10^5~\mathrm{cm}^{-3}$ (black). Temperature is shown on a
    positive linear scale between $0$ (black) and $10^4~\mathrm{K}$
    (white).  For each time, a sequence of 8 slices in the $yz$ plane
    are shown at $x$ intervals of 0.25~parsec, each with size 4 by 4
    parsecs ($x$ offsets from the ionizing star position are indicated
    on the bottom axis). This figure is also available as a color mpeg
    animation in the electronic edition of the \textit{Astrophysical
      Journal}. In the animation, the density of cold neutral gas is
    shown in positive grayscale, warm neutral gas in blue/green, and
    ionized gas in red/orange for three perpendicular midplane
    slices. }
  \label{fig:cuts}
\end{figure*}
The evolution of the ionized region is illustrated in
Figure~\ref{fig:cuts} by means of density and temperature slices in
different $yz$ planes for a sequence of times. At $t = 50,000$~years,
the ionized gas is confined to a small volume around the star, as can
best be appreciated in the temperature slices. One also sees high
temperature gas behind the shock that precedes the ionization front,
and which has now detached from the ionization front as it encounters
lower density gas. Between the shock and the ionization front is a
high density swept-up shell of neutral gas. By $t = 100,000$~years,
the ionized region has expanded considerably, reaching a radius of
$\sim 1$~parsec in most directions and even escaping from the grid in
one direction. Some dense neutral gas remains close to the ionizing
star, however, which forms photoablated globules and filaments. The
transonic flows from these can be seen to evacuate cavities in the
ionized gas, bounded by shocks. At later times, the photoablation
flows become even more prominent as more dense neutral clumps are
uncovered, while existing globules are eroded and pushed away from the
star by the rocket effect. By $t = 400,000$~years the ionization front
has escaped from the grid in most directions but some neutral gas
remains as close as 1~parsec to the ionizing star. The temperature is
roughly uniform in the ionized region at $\simeq 8000$~K, reaching
slightly higher values ($\simeq 10,000$~K) close to the ionization
front and in shocks, and slightly lower values ($\simeq 7000$~K) due
to expansion cooling in the photoablation flows.
\begin{figure}\centering
\includegraphics{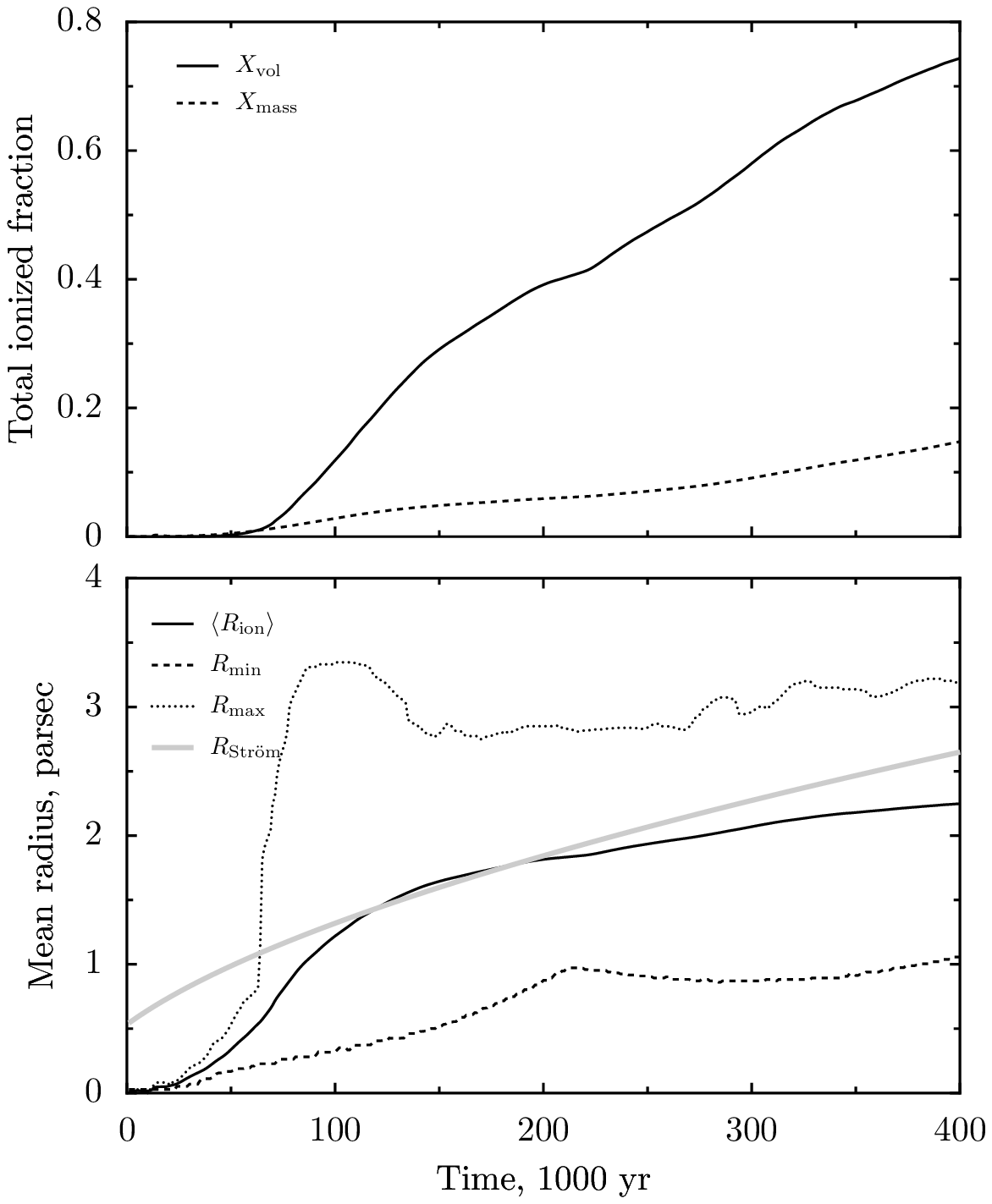}
\caption{Top: volume-weighted (solid line) and mass-weighted (dashed
  line) ionized fraction of the computational grid against
  time. Bottom: evolution with time of the mean radius (solid line) of
  the ionized region, compared with the minimum (dashed line) and
  maximum (dotted line) radius of the ionization front at each
  time. Also shown is the analytic solution (thick gray line) for the
  evolution of the mean radius of a Str\"omgren sphere in a
  homogeneous medium with the same mean density as our simulation.}
\label{fig:ionfrac}
\end{figure}

Figure~\ref{fig:ionfrac} (top panel) shows how the ionized fractions
of the total volume, $X_\mathrm{vol}$, and mass, $X_\mathrm{mass}$, of
our simulation varies with time. Both fractions are very small for the
first $\simeq 50,000$~years of evolution, during which the ionization
front remains trapped inside the dense clump in which the star formed.
By the end of our simulations, over 70\% of the volume of our cube has
been ionized, but only 15\% of the mass. However, this latter figure
is only a lower limit, since it only accounts for the ionized gas that
remains on the grid (see below). 

The lower panel of Figure~\ref{fig:ionfrac} shows the evolution of the
mean radius of the ionization front, averaged over all directions from
the ionizing source, $\langle R_\mathrm{ion} \rangle = ( 3
X_\mathrm{vol} V_\mathrm{grid} / 4 \pi )^{1/3}$, together with the
minimum and maximum ionization front radii at each time,
$R_\mathrm{min}$ and $R_\mathrm{max}$, respectively. The ratio
$R_\mathrm{max}/R_\mathrm{min}$ increases with time, indicating an
increasingly irregular shape for the ionized region, until at $t
\simeq 80,000$~years the ionization front suddenly escapes from our
computational box in some directions, at which point $R_\mathrm{max}$
becomes limited by the size of the grid. $R_\mathrm{min}$ remains
small at $\lesssim 1$~parsec during the entire simulation due to the
survival of dense neutral clumps as discussed above.  For comparison,
we also show the evolution of a D-type ionization front in a uniform
density medium with the same mean density as our simulations, which is
given by the expression \citep[e.g.,][]{1968dms..book.....S} $ R = R_0
( 1 + \frac74 c_\mathrm{i} t / R_0 )^{4/7}$ , where $c_\mathrm{i}$ is
the ionized sound speed and $R_0 = (S_* / \frac43 \pi n^2
\alpha_\mathrm{B})^{1/3} = 0.532$~parsec is the initial Str\"omgren
radius (the initial R-type propagation of the front lasts for only
$\simeq 100$~years for our parameters). At early times, our radii are
far below this prediction since the 1000 times overdensity at the
position of the ionizing star gives an initial Str\"omgren radius that
is 100 times smaller. By 150,000~years, the mean radius has caught up
with the uniform case, at which point its expansion appears to slow,
although this is probably simply an effect of the finite size of our
grid. The minimum radius is always well below the uniform case and
this is insensitive to the size of the grid.
\begin{figure}\centering
  \includegraphics{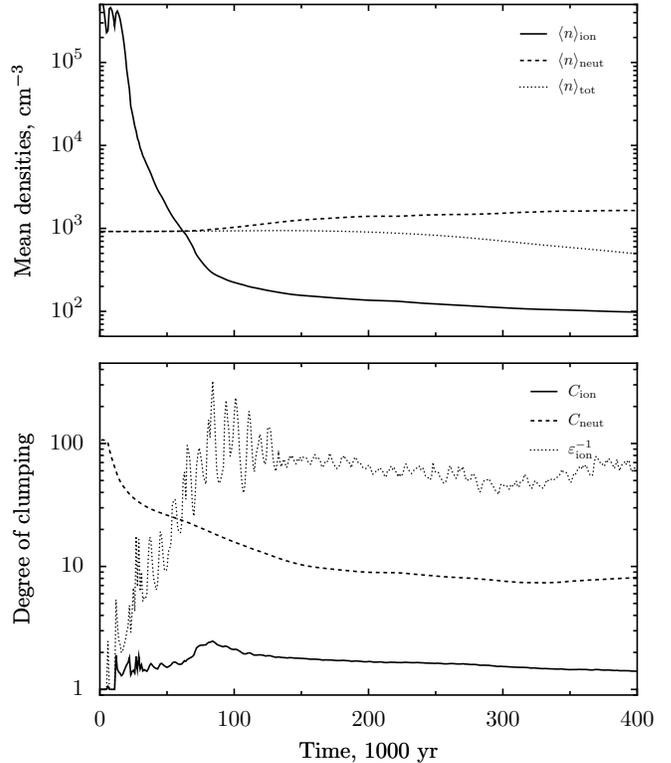}
  \caption{Top: Mean density, $\langle n \rangle$ against
    time. Bottom: Clumping, $C = \langle n^2 \rangle / \langle n
    \rangle^2$, against time. Solid line --- ionized gas; dashed line
    --- neutral gas; dotted line in bottom panel --- reciprocal of
    filling factor, $\varepsilon = \langle n^2 \rangle^3 / \langle n^3
    \rangle^2$, of ionized gas.}
\label{fig:clumping}
\end{figure}

Figure~\ref{fig:clumping} shows the evolution of the density in our
simulation. The mean density (top panel) in the ionized gas shows a
dramatic decline as the \Hii{} region expands, whereas the mean
density of the neutral gas shows a slight increase after 100,000~years
due to the effects of shocks driven by the expanding ionized
region. The slight decline in the mean total density towards the end
of the simulation indicates that we are starting to lose mass from our
grid. Although we do not track the ionization state of this gas, we
expect it to be predominantly ionized, giving an upper limit to the
fraction of the \emph{initial} simulation mass that has been ionized
by the end of the run of $\simeq 50\%$, as opposed to the lower limit
of 15\% that was derived above.  The density clumping of the neutral
gas (bottom panel) is initially very high due to the gravitational
collapse of dense regions in the turbulence simulations but quickly
falls with time as the densest region becomes photoionized and
expands. The clumping of the ionized gas is much less pronounced, with
$\langle n^2\rangle/\langle n \rangle^2$ not exceeding 2. However,
higher order statistics indicate significant inhomogeneities in the
ionized gas too, with a filling factor of $\varepsilon \sim 0.02$
during much of the evolution. This is primarily due to photoablation
flows from globules, which have ionized density contrasts of 3--10
with respect to the mean, and partly due to weak shocks, which have
typical density contrasts of 2--4.
\begin{figure}
  \centering
  \includegraphics{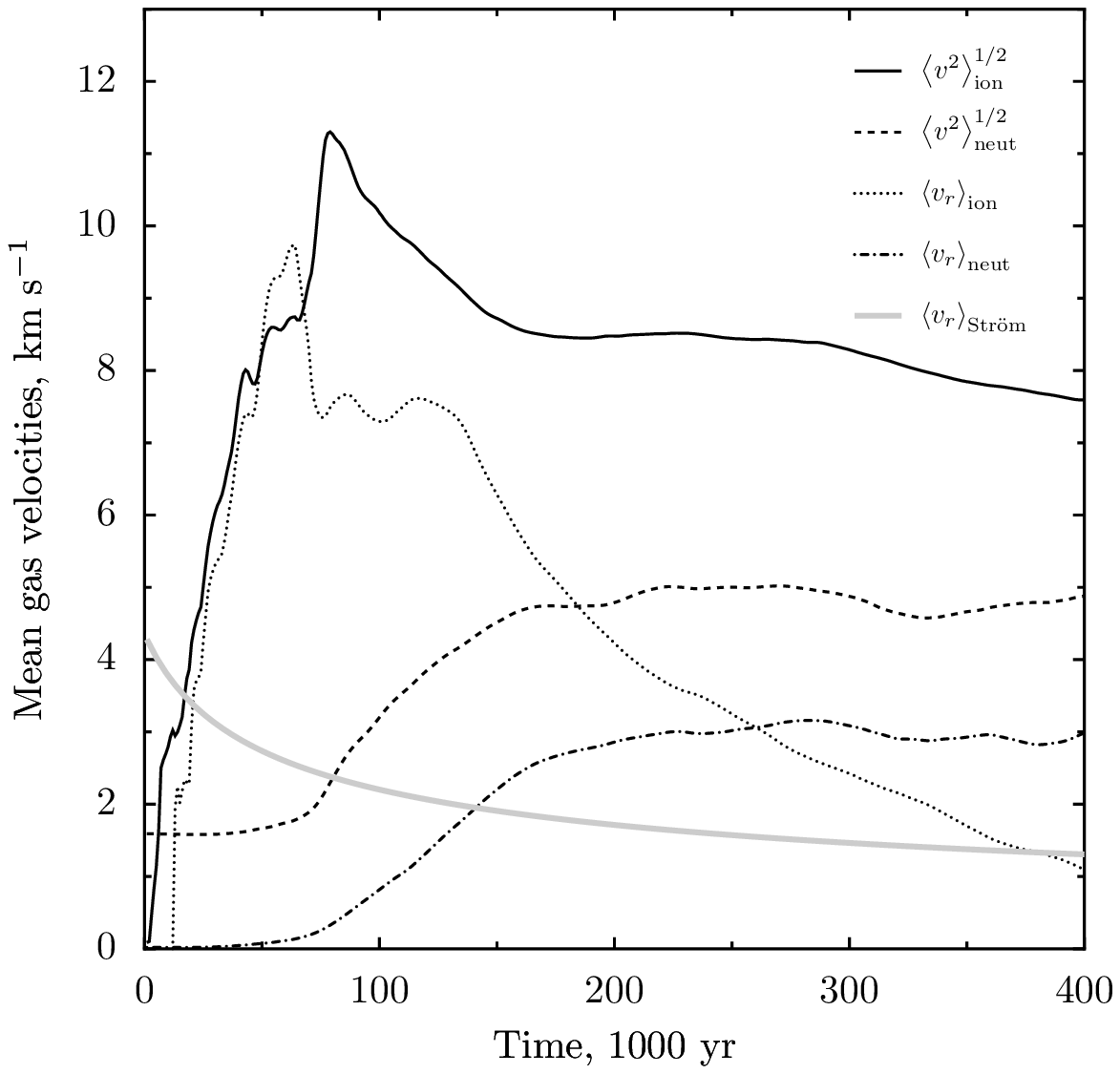}
  \caption{Average gas velocities against time. Solid line ---
    one-dimensional rms velocity of ionized gas; dashed line ---
    one-dimensional rms velocity of neutral gas; dotted line --- mean
    expansion velocity of ionized gas; dot-dashed line --- mean
    expansion velocity of neutral gas; thick gray line --- mean
    expansion velocity of ionized gas in a Str\"omgren sphere evolving
    in a uniform medium with the same mean density as our simulation.}
  \label{fig:vels}
\end{figure}

Figure~\ref{fig:vels} shows the mean gas velocities in the
simulations. The one-dimensional rms velocity over the volume, $V$, of
our simulation is defined as $$\langle v^2 \rangle^{1/2} = \left(\int\!
  \xi \left\vert \mathbf{v} \right\vert^2 d V \,\bigg/ \,3\!\! \int\! \xi\,
  dV\right)^{1/2},$$ and the mean expansion velocity as $$\langle v_r
\rangle = \int\! \xi \mathbf{v}.\mathbf{\hat{r}} \,d V \,\bigg/ \int\!
\xi\, dV,$$ where $\mathbf{v}$ is the vector gas velocity at a point,
$\mathbf{\hat{r}}$ is the unit vector in the radial direction away
from the ionizing star, and $\xi$ is the ionized or neutral fraction,
as appropriate. The rms velocity of the ionized gas increases with
time for the first 50,000~years until it is close to the ionized sound
speed, from which point on it remains roughly constant. The rms
velocity of the neutral gas remains at its initial value of $\sim
2~\kms$ for the first 100,000~years, after which it is increased to
$\simeq 5~\kms$ by the action of shocks driven by the expanding
ionized gas. The mean neutral expansion velocity is a significant
fraction of the rms velocity, indicating that the neutral gas motions
are predominantly away from the ionizing star. The same is true for
initial expansion of the ionized gas, but after 100,000 years the
ionized expansion velocity gradually declines to only $2~\kms$, while
the rms velocity remains at $8~\kms$. This is due to the effect of the
photoablation flows from dense globules, which flow back towards the
star at speeds of up to 20~\kms{} and partially cancel out the general
expansion of the ionized gas. The mean expansion velocity of the
ionized gas in a uniform medium is equal to $\frac38 d R / d t$ and is
always less than half of the ionized sound speed, which is much lower
than is seen in our simulation.

\newcommand\RedLine{[\ion{N}{2}] 6584~\AA\@}
\newcommand\GreenLine{H$\alpha$ 6563~\AA\@}
\newcommand\BlueLine{\ion{O}{3} 5007~\AA\@}
\begin{figure*}
  \setkeys{Gin}{width=0.32\linewidth}
  \includegraphics{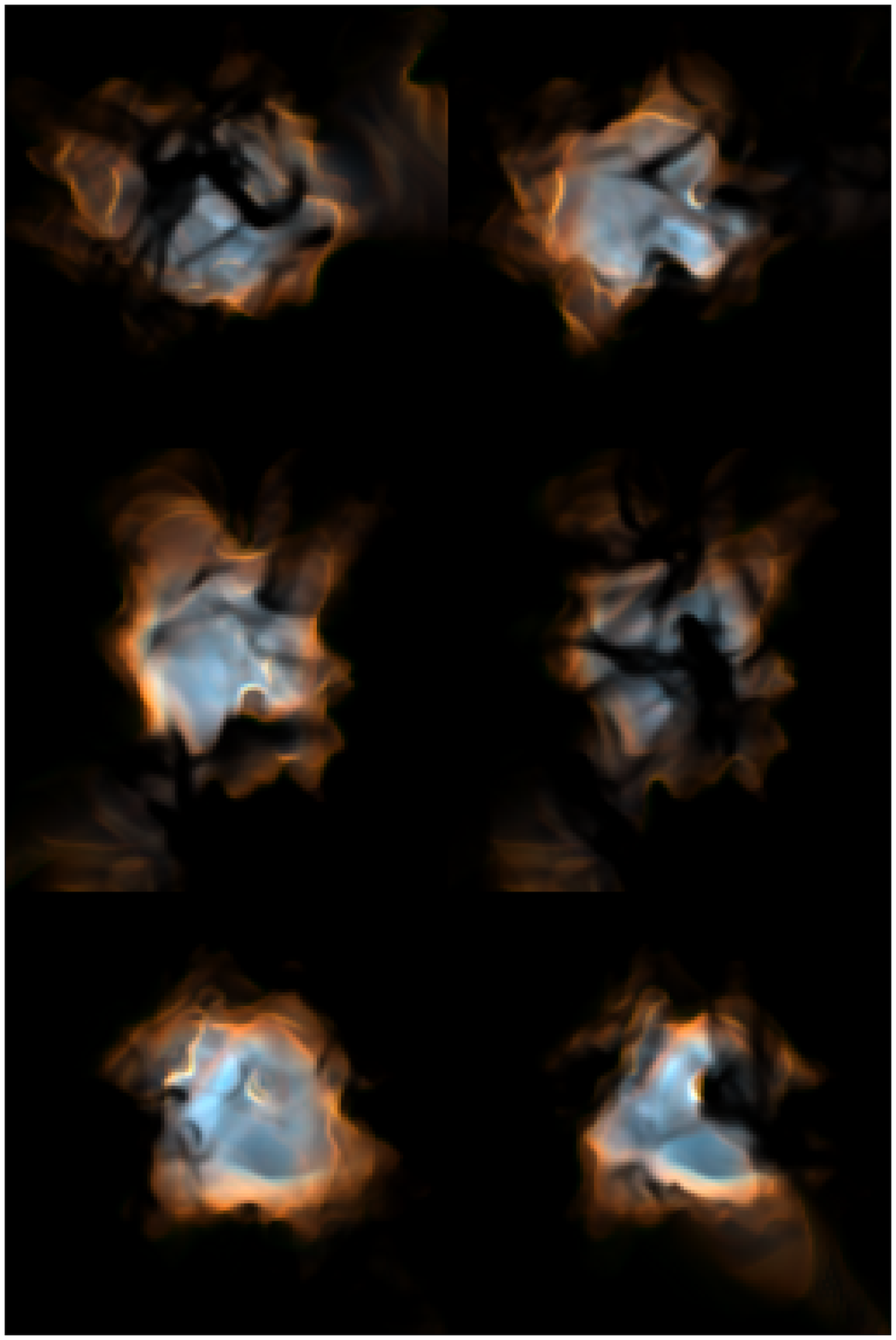}\hfill\includegraphics{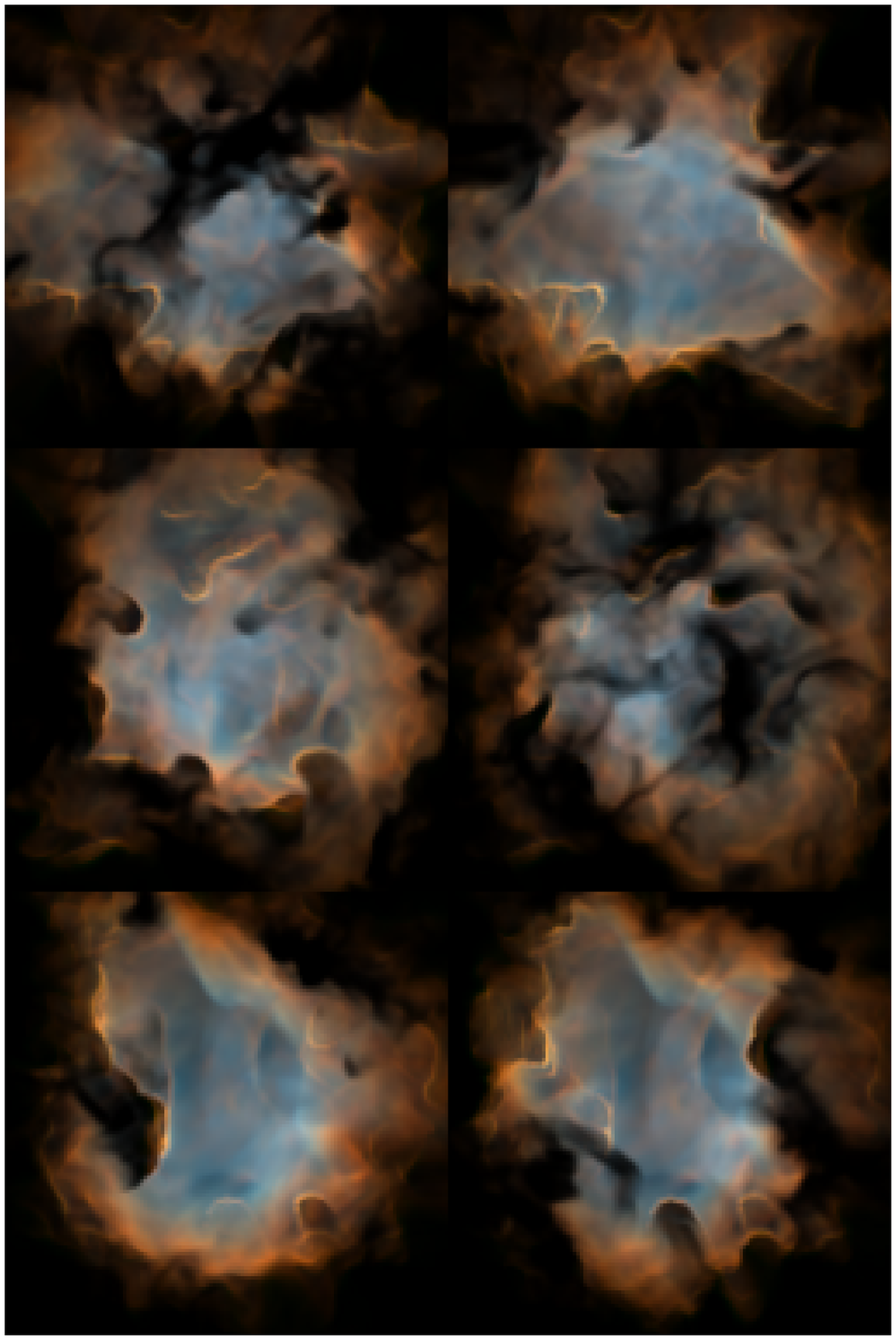}\hfill\includegraphics{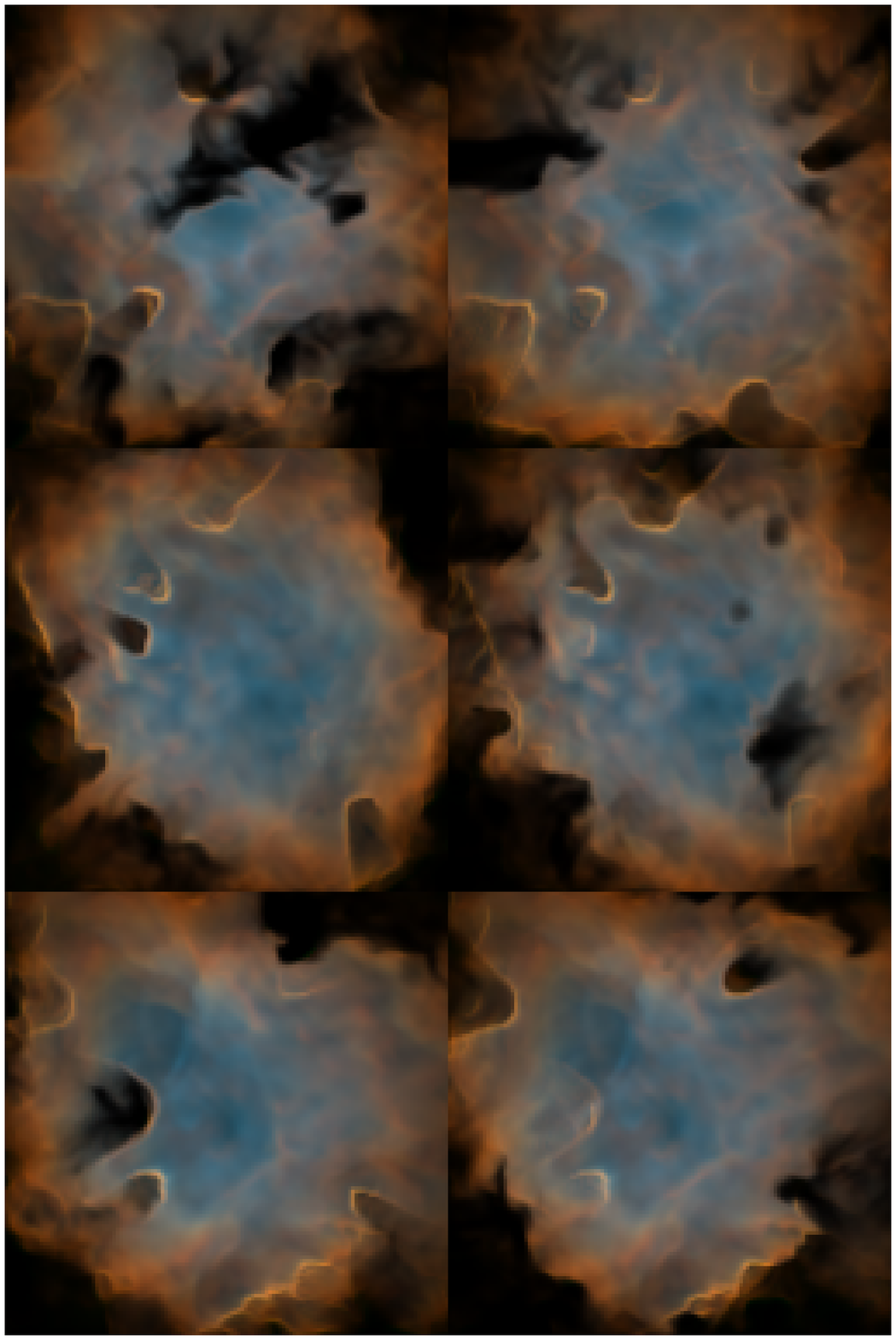}
  \caption[]{Synthetic narrow-band optical emission line images in the
    light of \RedLine{} (red), \GreenLine{} (green), and \BlueLine{}
    (blue) for our simulation at evolutionary times of (left
    panel)~100,000~years, (center panel)~250,000~years, and (right
    panel)~400,000~years.  In each case, projections are along the six
    principal axes of the simulation---top row: $+x$, $-x$; middle
    row: $+y$, $-y$; bottom row: $+z$, $-z$. The scales are linear in
    surface brightness, with maximum values of 0.005 (\RedLine{}),
    0.028 (\GreenLine{}), and 0.018 (\BlueLine{}), all in units of
    $\mathrm{erg\ s^{-1}\ cm^{-2}\ sr^{-1}}$ and assuming gas-phase
    abundances of $\mathrm{O/H} = 3 \times 10^{-4}$ and $\mathrm{N/H}
    = 5 \times 10^{-5}$.  }
  \label{fig:sim-emiss}
\end{figure*}
Figure~\ref{fig:sim-emiss} shows 3-color optical emission line images
of our simulations at three different evolutionary times, calculated
using 2-level atom approximations for the line emissivity as described
in \citet{2005ApJ...627..813H}. The ion fractions of N and O were
approximated as functions of the H ionization fraction, which were
calibrated using Cloudy \citep{2000RMxAC...9..153F}. This is a
reasonable approximation since departures from equilibrium ionization
should be similar for all ions \citep{2005ApJ...621..328H}.  The
images include the effects of local absorption by dust in the neutral
gas, with an assumed cross section of $5\times 10^{-22}~\mathrm{cm^2}$
per H nucleon \citep{1991ApJ...374..580B} but do not include
scattering. They are designed to be directly comparable with observed
narrow-band emission line observations of real nebulae. In panel
\textit{a}, although the ionization front has already broken out in
some directions, this is only where the density is very low, resulting
in weak emission. Bright cusps indicate the ionization fronts at the
heads of photoablated globules that protrude into the interior of the
ionized region. On a larger scale, the outer border of the emission
seems to consist of several roughly straight segments. This is due to
the formation of cylindrical ionization fronts around filaments in the
neutral density distribution.  Dark lanes are due to dust absorption
in dense filaments that lie outside the ionized region. As the
simulation progresses (panels \textit{b} and \textit{c}), one sees a
greater number of photoablated globules and at the same time, the
overlying extinction becomes less. Shocks due to the collisions
between photoablation flows produce brightness variations in the fully
ionized gas.

\section{Discussion and conclusions}
\label{sec:discuss}

\Hii{} regions around single stars are found with a variety of sizes,
from ultracompact ($<0.01$~parsec) to extended ($> 1$~parsec) and it
is probable that these represent an evolutionary sequence
\citep[e.g.,][]{1996ApJ...469..171G}. Our simulation passes through
all of these phases during its evolution but, because of the physical
scaling we have chosen, the initial compact phases are very poorly
resolved. We therefore compare our results with the observed structure
and dynamics of more evolved, extended \Hii{} regions. The shapes of
such regions tend to be very irregular on a large scale and frequently
show bright-rimmed structures on smaller scales
\citep{1956BAN....13...77P}, together with filamentary overlying
extinction features. Many Galactic \Hii{} regions show strong
similarities in appearance with our simulation (see especially
Fig.~\ref{fig:sim-emiss}), with perhaps the most striking resemblance
occuring in the case of M20, the Trifid Nebula
\citep[see][]{2005sfet.confE..15R}. In the case of many other regions,
such as the Orion Nebula (M42), and the Eagle Nebula (M16), the
details of the observed emission structures are very similar to those
seen in our simulations but the large-scale distributions are somewhat
different. This is presumably due to density gradients on a scale
equal to or larger than that of the \Hii{} region itself, which give
rise to champagne flows that are ionization-bounded on one side
\citep{1979ApJ...233...85B, 1979A&A....71...59T, 2005ApJ...627..813H,
  2005astro.ph.11035A}. These are not present on our simulation, in
which the density is roughly homogeneous on scales $\gtrsim 1$~parsec,
although the effects of such large-scale gradients can be clearly seen
in the simulations of \citet{2005MNRAS.358..291D}. On an even larger
scale, very similar emission structures are seen in giant \Hii{}
regions, of which NGC~3603 in our Galaxy \citep{2002ApJ...573..191M}
and 30~Doradus in the Large Magellanic Cloud
\citep{1998AJ....116..163S} are the best-studied examples. However,
such regions are ionized by multiple stars and therefore cannot be
directly compared with our simulations, although similar processes are
expected to occur \citep{2006astro.ph..1631T}.

There are hints in our simulation of a qualitative change in the
morphology of the region as it evolves. At times soon after the
breakout of the ionization front from the dense core, one finds
photoablating structures that are predominantly cylindrical in form,
oriented side-on to the flux of ionizing photons. These give rise to
bar-like features in the emission maps, which are similar in
appearance to the Bright Bar and other similar features found in the
Orion Nebula \citep{2000AJ....120..382O}. At later times, the
photoablating structures become more finger-like, with a spherical
ablation flow from the bright tip of the finger, which points towards
the ionizing star. An additional feature of the late-time evolution of
our simulation is the formation of recombination fronts, which
temporarily reverse the expansion of the ionized region in certain
directions. These form due to collisions between ionized flows,
forming dense shells that in certain circumstances are capable of
trapping the ionization front. Although these recombination fronts
also show finger-like irregularities, they are easily distinguished
from the photoablating fingers because the ionized density does not
show a maximum at the ionization front. As a result, they give soft,
diffuse edges to the \Hii{} region in the emission maps, as opposed to
the sharp edges associated with the photoablation flows.

The rms velocity of the ionized gas is remarkably constant at $\simeq
8~\kms$ after the initial breakout of the ionization front from the
dense core, which occurs around $t \simeq 50,000$~years. For the first
$\simeq 150,000$~years, this is driven largely by the physical
expansion of the ionized core in a mode similar to the fully
density-bounded champagne flows studied by
\citet{2002ApJ...580..969S}. At later times, it is the divergent
photoablation flows from numerous neutral globules that maintain the
rms velocity. The rms velocity of the neutral gas is also strongly
affected by the ionized region, increasing to $\simeq 5~\kms$ after
150,000~years, which is 2--3 times the initial turbulent
value. Optical spectroscopy of \Hii{} regions frequently shows
non-thermal linewidths of $\sim 10~\kms{}$, which are roughly constant
between partially ionized and fully-ionized species
\cite{2001ARA&A..39...99O, 2003AJ....125.2590O}. This is difficult to
explain in the context of \Hii{} regions in smooth density gradients
\citep{2005ApJ...627..813H} but would be a natural consequence of our
simulation.

Our simulations are only a first step towards fully self-consistent
models of \Hii{} region evolution in realistic density
distributions. Important physical processes that we have neglected
include the stellar wind from the ionizing star and the radiative
acceleration of dust grains, which are coupled to the gas via Coulomb
collisions. Both these would tend to evacuate a central cavity in the
photoionized gas. Even in smooth density distributions the effect of
stellar winds on the dynamics of the ionized gas is found to be rather
complex \citep{2005astro.ph.11035A}. The effects of magnetic fields on
the ionization fronts and on the dynamics of the neutral gas
\citep{1998A&A...331.1099R, 2000MNRAS.314..315W, 2001MNRAS.325..293W}
is another area that needs to be explored, since it is possible that
MHD effects partially dictate the shape of the photoablated structures
\citep{2003A&A...403..399C, 2005Ap&SS.298..183R, 2006-FS-Robin}.  A
more realistic treatment of the diffuse ionizing field may also affect
the details of the shapes of the photoablated globules but should not
significantly change the dynamics. The evolution of the region on
timescales longer than the 400,000~year duration of our simulation is
also of interest, but would require the consideration of a larger
computational domain.

In summary, we have carried out the first investigation of the
hydrodynamic evolution of an \Hii{} region in a realistic turbulent
medium. We have shown that the clumpy and filamentary structure of the
underlying density distribution leads to an extremely irregular shape
for the ionization front. The density structure in the ionized region
is less clumped than that of the neutral gas, but nevertheless shows
complex dynamical structures due to the mutual interactions between
photoablation flows from neutral globules and filaments, some of which
survive at close distances to the ionizing star ($< 1$~parsec)
throughout the evolution. The rms velocity of the ionized gas is
approximately equal to the ionized sound speed ($\sim 10~\kms$) during
almost the entire evolution, even at late times when the net radial
expansion is very low ($< 2~\kms$). The calculated emission-line
morphologies of our simulation show striking similarities to the
appearance of real \Hii{} regions.

\acknowledgments We are indebted to Enrique V\'azquez-Semadeni for
generously providing us with the results of the numerical turbulence
simulations used as initial conditions in this paper. GM acknowledges
financial support from the Royal Netherlands Academy of Arts and
Sciences and is grateful for the financial support from the
Universidad Nacional Aut\'onoma de M\'exico, which made possible a
visit to the CRyA in Morelia during which most of the work for this
paper was carried out.  SJA and WJH acknowledge support from
DGAPA-UNAM, through project PAPIIT 112006-3. PRS is grateful for the
support of NASA Astrophysical Theory Program Grant NNG04G177G\@. We
thank the anonymous referee for a helpful report. This work has made
extensive use of NASA's Astrophysics Abstract Data Service and the
astro-ph archive.


\end{document}